\documentclass[12pt]{article}
\usepackage{amscd,amssymb,amsmath,latexsym,enumerate}
\usepackage[mathscr]{euscript}
\usepackage{epsfig}
\usepackage{fancybox}
\usepackage{verbatim}

\usepackage{color}
\newcommand{\red}{\textcolor[rgb]{0.50,0.00,0.00}} 
\title{Spectral flows of dilations of Fredholm operators}

\author{Giuseppe De Nittis, Hermann Schulz-Baldes
\\
\\
{\small Department Mathematik, Universit\"at Erlangen-N\"urnberg, Germany}
}

\date{ }

\newtheorem{theo}{Theorem}
\newtheorem{defini}{Definition}
\newtheorem{proposi}{Proposition}
\newtheorem{lemma}{Lemma}

\newcommand{\BM}{{\mathbb B}}
\newcommand{\CM}{{\mathbb C}}
\newcommand{\EM}{{\mathbb E}}
\newcommand{\NM}{{\mathbb N}}
\newcommand{\RM}{{\mathbb R}}
\newcommand{\SM}{{\mathbb S}}

\newcommand{\ZM}{{\mathbb Z}}
\newcommand{\PM}{{\mathbb P}}

\newcommand{\KM}{{\mathbb K}}

\newcommand{\UM}{{\mathbb U}}

\newcommand{\FM}{{\mathbb F}}

\newcommand{\Aa}{{\cal A}}

\newcommand{\Dd}{{\cal D}}

\newcommand{\Vv}{{\cal V}}

\newcommand{\Tr}{\mbox{\rm Tr}}

\newcommand{\Mm}{{\cal M}}
\newcommand{\Cc}{{\cal C}}

\newcommand{\Ll}{{\cal L}}

\newcommand{\Kk}{{\cal K}}
\newcommand{\Hh}{{\cal H}}

\newcommand{\one}{{\bf 1}}

\newcommand{\ev}{{\mbox{\rm ev}}}
\newcommand{\SF}{{\rm Sf}} 
\newcommand{\Ind}{{\rm Ind}} 
\newcommand{\Ker}{{\rm Ker}} 
\newcommand{\Ran}{{\rm Ran}}

\newcommand{\ess}{{\mbox{\rm\tiny ess}}}

\begin{document}

\maketitle

\begin{abstract}
Given an essentially unitary contraction and an arbitrary unitary dilation of it, there is a naturally associated spectral flow which is shown to be equal to the index of the operator.  This result is interpreted in terms of the $K$-theory of an associated mapping cone. It is then extended to connect $\ZM_2$ indices of odd symmetric Fredholm operators to a $\ZM_2$-valued spectral flow.
\end{abstract}

\vspace{1cm}


The spectral flow of a one-parameter family of self-adjoint Fredholm operators was introduced by Atiyah, Patodi and Singer \cite{APS} and was under certain conditions shown to be connected to  indices of Fredholm operators. A particularly convenient technical reformulation was given by Phillips \cite{Phi1} which also extends to unbounded operators \cite{CP}. As this version of spectral flow is also used here, it is reviewed in Section~\ref{sec-specflow} below. The first part of the paper then considers an essentially unitary contraction operator $T$ on a Hilbert space $\Kk$ and an arbitrary unitary dilation $U_T$ associated to it, namely a unitary operator on a larger Hilbert space $\Hh$ so that its compression to $\Kk$ is $T$. If $P$ is the projection onto $\Kk$ and one sets $F=2P-\one$, then Theorem~\ref{theo-FredFlow} in Section~\ref{sec-specflowFred} shows that the spectral flow of any path $s\in[0,1]\mapsto F_s$ from $F$ to $U_T^*FU_T$ and with compact $F_s-F$ is equal to the index of $T$, up to a sign. Two proofs are provided, one by a homotopy argument and one using indices of pairs of projections \cite{Kat,ASS}. This result already appears in \cite{Phi2}, but for sake of completeness and as preparation for the second part of the paper, we provide two proofs of it. Inspired by \cite{Put,CPR}, Section~\ref{sec-mappingcone} then provides a $K$-theoretic interpretation of the theme {\it index = spectral flow} based on a mapping cone exact sequence and the pairing of its $K$-groups with adequate Fredholm modules. Indeed, the exposition makes explicit the tight connection between unitary dilations and Fredholm modules \cite{HR}.

\vspace{.2cm}

More novel is an extension of the above result to essential untaries operators and dilations having symmetry properties linked to a real or quaternionic structure on the Hilbert space. Examples are real, quaterionic, symmetric and anti-symmetric operators. Anti-symmetric operators are in bijection with so-called odd symmetric operators and such Fredholm operators have recently been shown to have $\ZM_2$ indices given by the dimension of their kernel modulo $2$ \cite{Sch}. All this is reviewed in Section~\ref{sec-struct}. Let us also note that these $\ZM_2$ indices are defined for a much wider class of Fredholm operators than considered by  Atiyah and Singer \cite{AS} which required the anti-symmetric Fredholm operators also to be real. In Section~\ref{sec-Z2flow} a $\ZM_2$-valued spectral flow is defined for paths with an adequate symmetry property. The basic idea leading to this definition is already contained in the analysis of $\ZM_2$ invariants of edge states in quantum systems with odd time-reversal symmetry \cite{ASV}. In Section~\ref{sec-oddsympSF} this $\ZM_2$-valued spectral flow of an odd symmetric dilation is again shown to be equal to the $\ZM_2$ index of the dilated Fredholm operator. A mapping cone interpretation with the flavor of Real $K$-theory \cite{Sc} is sketched in the final Section~\ref{sec-mappingconeodd}.

\vspace{.2cm}

Our motivation to revisit the problem of spectral flow and to extend it to $\ZM_2$ indices stems from applications to solid state physics. This will be the subject of a companion paper \cite{DS}.

\vspace{.3cm}

\noindent {\bf Acknowledgement:} We thank Magnus Goffeng for correcting several mistakes in a first version of the manuscript. We received financial support of the DFG and the Alexander von Humboldt Foundation.

\section{Spectral flow of a pair of unitary equivalent operators}
\label{sec-specflow}

In order to fix notations and terminology, this section reviews some standard facts from the theory of Fredholm operators on a separable Hilbert space $\Hh$ as well as the notion of spectral flow. Let $\BM(\Hh)$, $\KM(\Hh)$ and $\FM(\Hh)$ denote the sets of bounded, compact and Fredholm operators on $\Hh$, respectively. Recall that  $F\in\BM(\Hh)$ is Fredholm if and only if it has closed range $T\Hh$ and finite-dimensional kernel $\Ker(F)$ and cokernel $\Ker(F^*)$. Atkinson's theorem states that Fredholm operators are the invertibles modulo compact perturbation and this provides the following characterization 
$$
\FM(\Hh)
\;=\;
\big\{F\in\BM(\Hh)\;|\; \exists\;S\in\BM(\Hh),\;\;K_1,K_2\in\KM(\Hh)\;\;\text{with}\;\;SF-K_1=\one=FS-K_2\big\}\;.
$$
In particular, the property to be Fredholm is invariant under compact perturbations. Furthermore, $\FM(\Hh)$ is an open subset of $\BM(\Hh)$ which is closed under the adjoint involution. The Noether index of a Fredholm operator $F$ is defined by $\Ind(F)=\dim(\Ker(F))-\dim(\Ker(F^*))\in\ZM$. The index is a homotopy invariant that is stable under compact perturbations, namely $\Ind(F)=\Ind(F+K)$ for all $K\in\KM(\Hh)$. Moreover, $\FM_n(\Hh)=\Ind^{-1}(n)$ is a path connected component for any $n\in\ZM$ so that the index map $\Ind:\FM(\Hh)\to \ZM$ establishes a bijection between $\ZM$ and the path-connected components of $\FM(\Hh)=\bigcup_{n\in\ZM}\FM_n(\Hh)$. The set $\FM(\Hh)$ has a group structure under the multiplication of operators and the  index map is a group homomorphism onto the additive group $\ZM$. 

\vspace{.2cm}

Let $\SM\FM(\Hh)=\{F\in\FM(\Hh)\;|\; F=F^*\}$ be the subset of self-adjoint Fredholm operators. Equivalently,  $\SM\FM(\Hh)=\{F\in\BM(\Hh)\;|\; F=F^*\mbox{ and }0\not\in\sigma_\ess(F)\}$ where $\sigma_\ess(F)$ denotes the essential spectrum. Although $\SM\FM(\Hh)\subset \FM_0(\Hh)$, the space $\SM\FM(\Hh)$ has a non-trivial topology. First of all, it has three disjoint components:
$$
\SM\FM_\pm(\Hh)\;=\;\big\{F\in \SM\FM(\Hh)\;|\;\sigma_{\rm ess}(F)\subset\RM_\pm\big\}\;,
\qquad
\SM\FM_*(\Hh)\;=\;\big\{F\in \SM\FM(\Hh)\;|\;\sigma_{\rm ess}(F)\cap\RM_\pm\neq\emptyset\big\}
\;,
$$
where the notation $\RM_\pm=\{x\in\RM\,|\,\pm x>0\}$ was used. The two components $\SM\FM_\pm(\Hh)$ are contractible \cite[Theorem B]{AS}, but $\pi_1\big(\SM\FM_*(\Hh)\big)\cong\ZM$ via the spectral flow isomorphism. The spectral flow was introduced in \cite{APS} using the intuitive notion of intersection theory of spectral curves. Here we rather work with the versatile, but equivalent notion of spectral flow proposed in \cite{Phi1}. 

\begin{figure}
\begin{center}
\includegraphics[height=7cm]{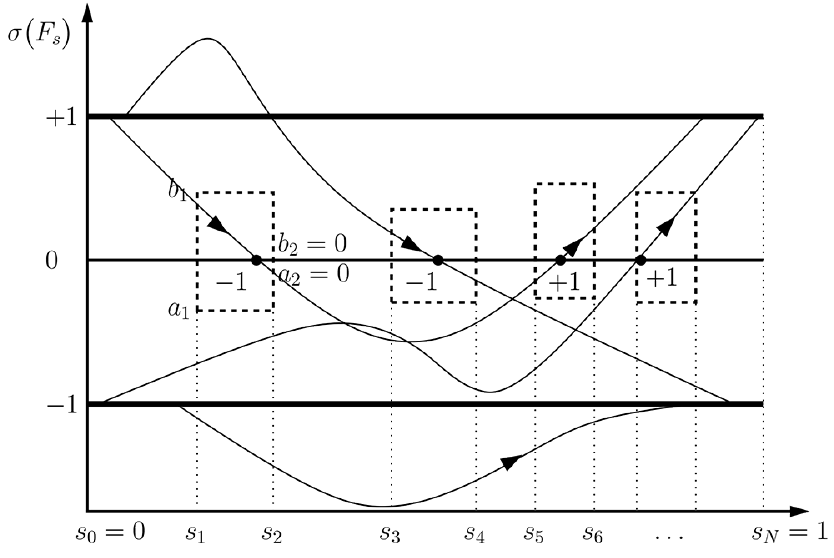}
\caption{\it Schematic representation of the objects used in the definition {\rm \eqref{eq-SFdef}} of the spectral flow as well as in the second proof of {\rm Theorem~\ref{theo-FredFlow}}. Far from the crossings it is possible to set $a=b=0$.}
\end{center}
\end{figure}

\vspace{.2cm}

Let $s\in [0,1]\mapsto F_s\in \SM\FM_*(\Hh)$ be a continuous path, not necessarily closed. For $a\in(-1,0]$ and $b\in[0,+1)$ set
$$
Q_{a,b}(s)\;=\;
\chi_{(a,b]}(F_s)
\;,
$$
where $\chi_I$ denotes the characteristic function on $I\subset\RM$. By compactness (see Figure~1 where $\sigma_\ess(F)=\{-1,1\}$ is assumed), it is possible to choose a finite partition $0=s_0<s_1<\ldots<s_{N-1}<s_N=1$ of $[0,1]$ and $a_n< 0 < b_n$, $n=1,\ldots,N$, such that $s\in [s_{n-1},s_n]\mapsto Q_{a_n,b_n}(s)$ is continuous with  constant (necessarily) finite  rank. Then define the spectral flow by
\begin{equation}
\label{eq-SFdef}
\SF(s\in[0,1]\mapsto F_s)
\;=\;
\sum_{n=1}^N {\rm Tr}_{\Hh}\left(Q_{a_n,0}(s_{n-1})\;-\; Q_{a_n,0}(s_n)\right)
\;.
\end{equation}
Note that $Q_{a_n,0}(s_{n-1})$ and $Q_{a_n,0}(s_n)$ are both finite dimensional projections so that the trace is finite. The basic result about the spectral flow is that it is well-defined by the above procedure and it is homotopy invariant. A detailed proof can be found in \cite{Phi1}.

\begin{theo}
\label{theo-hmot_inv} 
The definition of $\SF(s\in[0,1]\mapsto F_s)$ is independent of the choice of the partition $0=s_0<s_1<\ldots<s_{N-1}<s_N=1$ of $[0,1]$ and values $a_n< 0 < b_n$ such that $s\in [s_{n-1},s_n]\mapsto Q_{a_n,b_n}(s)$ is continuous. Moreover, let $s\in [0,1]\mapsto F_s$ and $s\in[0,1]\mapsto G_s$ be two continuous paths in $\SM\FM_*(\Hh)$ such that $F_0=G_0$ and $F_1=G_1$. Then $\SF(s\in [0,1]\mapsto F_s)=\SF(s\in [0,1]\mapsto G_s)$ if and only if there exists a norm continuous homotopy between the two paths leaving the endpoints fixed.
\end{theo}

\noindent {\bf Remark} 
Let us briefly sketch the connection of definition \eqref{eq-SFdef} to the intuitive notion of spectral flow. By standard perturbation theory arguments \cite{Kat}, it is possible to label the spectral curves $\lambda_j(s)$ such that each varies continuously in $s$. When $s$ increases the spectral curves $s\mapsto \lambda_j(s)$ can cross the segment $[0,1]\times\{0\}$. One has a spectral crossing of positive signature if there is a passage  from a negative to a positive eigenvalue or a  spectral crossing of negative signature if the passage is from a  positive to a negative eigenvalue. If there is a finite number of crossings and no crossings at the boundaries $s=0$ and $s=1$, the sum of these signatures over all crossings is equal to $\SF(s\in[0,1]\mapsto F_s)$. The advantage of definition \eqref{eq-SFdef} is that the boundaries do not require special treatment and that there may well be an infinite number of crossings.
\hfill $\diamond$

\vspace{.2cm}

When Theorem~\ref{theo-hmot_inv} is applied to closed paths, it shows that $\SF:\pi_1\big(\SM\FM_*(\Hh)\big)\to\ZM$ is a well-defined group homomorphism. To verify that this map is bijective, one can use the fact that the space 
$$
\widehat{\SM\FM}_*(\Hh)\;=\;\big\{F\in \SM\FM_*(\Hh)\;|\;\|F\|=1\;, \;\; \sigma_{\rm ess}(F)=\{-1,1\}\big\}
$$ 
is a deformation retract of $\SM\FM_*(\Hh)$ and that the map ${\rm \varphi}:\widehat{\SM\FM}_*(\Hh)\to \UM_*(\Hh)$ defined by $\varphi(F)=e^{\imath\pi(F+\one)}$ is an homotopy equivalence \cite{AS}. Here  $\UM_*(\Hh)$ is the subgroup of those unitary  operators $U\in\UM(\Hh)$ for which $U-\one\in\KM(\Hh)$. The sequence of isomorphisms  $\pi_1\big(\SM\FM_*(\Hh)\big)\cong\pi_1\big(\widehat{\SM\FM}_*(\Hh)\big)\cong\pi_1\big(\UM_*(\Hh)\big)$ combined with the standard isomorphism $\pi_1\big(\UM_*(\Hh)\big)\to\ZM$ given by the \emph{winding number} concludes the description of the fundamental group of $\SM\FM_*(\Hh)$. Just for sake of completeness let us recall that the winding number of a differentiable loop $s\in [0,1] \mapsto U_s\in\UM_*(\Hh)$ is given by
$$
\mbox{\rm Wind}(s\in[0,1]\mapsto U_s)
\;=\;
\frac{1}{2\pi\imath}\int_0^1 ds\; \Tr\left((U_s)^{-1}\partial_sU_s\right)
\;,
$$
whenever $\partial_sU_s$ is traceclass. For adequate paths, the equality $\SF(s\in[0,1]\mapsto F_s)=\mbox{\rm Wind}(s\in[0,1]\mapsto \varphi(F_s))$ provides an alternative formula for the computation of the spectral flow.

\vspace{.2cm}

Focus here will be on the spectral flow of certain special paths in $\widehat{\SM\FM}_*(\Hh)$. Given $F\in\widehat{\SM\FM}_*(\Hh)$  and a unitary $U\in\UM(\Hh)$, let $\Theta(F,U)$ denote the set of paths $s\in[0,1]\mapsto F_s$ such that
\begin{align}
{\rm (i) } & \;\;\;\; F_0=F\;,
\nonumber
\\
{\rm (ii) } & \;\;\;\; F_s-F\in \KM(\Hh) \;\;\mbox{\rm for all } s\in[0,1]\;,
\label{eq-SFcond}
\\
{\rm (iii) } & \;\;\;\; F_1=U^* F U\;.
\nonumber
\end{align}
Note that  $\sigma(F_1)=\sigma(F_0)$. The spectral flow $\SF(s\in[0,1]\mapsto F_s)$ of such a path is well-defined and equal to the spectral flow of the path obtained by concatenation with the isospectral path $r\in[0,1]\mapsto (U^r)^*FU^r$. 

\begin{proposi}
\label{prop-SFFU} 
The spectral flow $\SF:\Theta(F,U)\to \ZM$ is equal to a constant denoted by $\SF(F,U)$. In particular,
\begin{equation}
\label{eq-SFFU}
\SF(F,U)\;=\; \SF(s\in[0,1]\mapsto F+s\;U^*[F,U])\;.
\end{equation}
\end{proposi}

\noindent {\bf Proof.} Let $s\in[0,1]\mapsto F_s$ and $s\in[0,1]\mapsto F'_s$ be two paths in $\Theta(F,U)$. Then $r\in[0,1]\mapsto rF_s+(1-r)F'_s$ is a homotopy in $\Theta(F,U)$ keeping the initial and final point fixed, so that Theorem~\ref{theo-hmot_inv} implies the first claim. The second follows because $s\in[0,1]\mapsto F+s\;U^*[F,U]$ is indeed a path in $\Theta(F,U)$ since $U^*[F,U]=F_1-F_0\in\KM(\Hh)$. 
\hfill $\Box$

\vspace{.2cm}

Let us introduce the set of operator pairs for which a spectral flow can be defined:
$$
\PM(\Hh)
\;=\;
\left\{
(F,U)\in\widehat{\SM\FM}_*(\Hh)\times \UM(\Hh)
\;\left|\;
[F,U]\;\mbox{\rm compact}
\right.
\right\}
\;.
$$
It carries the subspace topology induced from the norm topology on $\BM(\Hh)\times\BM(\Hh)$. Note that indeed each point $(F,U)\in\PM(\Hh)$ defines a class of paths $\Theta(F,U)$ so that it is possible to view the spectral flow as a map $\SF:\PM(\Hh)\to \ZM$ (strictly speaking this is $\SF\circ\Theta$).

\begin{proposi}
\label{prop-SFFU2} 
The spectral flow $\SF:\PM(\Hh)\to \ZM$ is locally constant.
\end{proposi}

\noindent {\bf Proof.}  Let $(F(0),U(0))$ and $(F(1),U(1))$ be two points connected by a continuous path $r\in[0,1]\mapsto(F(r),U(r))$ in $\PM(\Hh)$. This means that there is a path $\alpha$ in $\widehat{\SM\FM}_*(\Hh)$, given by $r\mapsto F(r)$,  which connects $F(0)$ with $F(1)$ and a second path  $\beta$ in $\widehat{\SM\FM}_*(\Hh)$, given by $r\mapsto U(r)^*F(r)U(r)$,  which connects $U(0)^*F(0)U(0)$ with $U(1)^*F(1)U(1)$. Let $\gamma_r\in\Theta(F(r),U(r))$ by any path connecting $F(r)$ to $U(r)^*F(r)U(r)$ by compact perturbations. For each $r\in[0,1]$ let us denote by $\alpha_r$ the reduced path  which connects $F(0)$ with $F(r)$ along the path $\alpha$. In similar way let us  introduce also the reduced path $\beta_r$. The composed paths $\theta_r=\beta_r^{-1}\circ\gamma_r\circ\alpha_r$ produce a homotopy between $\theta_0=\gamma_0$ and $\theta_1$ within the class of paths having same extreme points $F(0)$ and $U(0)^*F(0)U(0)$. Theorem~\ref{theo-hmot_inv} applies so that $\SF(\theta_1)=\SF(\theta_0)=\SF(F(0),U(0))$. Now, by applying the composition rule of the spectral flow $\SF(\theta_1)=\SF(\alpha)+\SF(\gamma_1)-\SF(\beta)$ and observing that $\SF(\alpha)=\SF(\beta)$ one gets $\SF(F(0),U(0))=\SF(\gamma_1)=\SF(F(1),U(1))$.
\hfill $\Box$

\section{Dilation of a Fredholm operator and its spectral flow}
\label{sec-specflowFred}

In this section, the index of an arbitrary Fredholm operator is being calculated as a spectral flow, actually of a whole family of spectral flows associated to arbitrary dilations. Let $\Kk$ be a separable Hilbert space.  Let $T\in\BM(\Kk)$ be a contraction, namely $\|T\|\leq 1$. A unitary dilation of  $T$ is a unitary operator $U_T\in\BM(\Hh)$ on some Hilbert space $\Hh$ in which $\Kk$ is isometrically embedded by an injective partial isometry $\Pi:\Kk\hookrightarrow\Hh$ such that
\begin{equation}
\label{eq-dildef}
T\;=\;\Pi^*\,U_T\,\Pi\;.
\end{equation}
To exclude a trivial case, it will always be assumed that $\Hh\ominus \Kk$ is infinite dimensional.  Recall that an operator $T$ is called essentially unitary if $T^*T-\one$ and $TT^*-\one$ are compact operators. The set $\EM\UM(\Kk)$ of essentially unitary operators is a subset of the  Fredholm operators $\FM(\Kk)$.

\begin{theo}
\label{theo-FredFlow} {\rm \cite{Phi2}} Let $U_T\in\UM(\Hh)$ be a unitary dilation of an essentially unitary contraction $T\in \EM\UM(\Kk)$ and let $\Pi:\Kk\hookrightarrow\Hh$ be the associated injective partial isometry. Then, with  $F=2\,\Pi\Pi^*-\one\in \widehat{\SM\FM}_*(\Hh)$ 
\begin{equation}
\label{eq-SFInd}
\Ind(T)\;=\;-\,\SF(F,U_T)\;=\;-\,\SF\bigl(s\in[0,1]\mapsto F+s\;U_T^*[F,U_T] \bigr)
\;.
\end{equation}
\end{theo}

Two proofs will be provided, one by constructing a homotopy to a special dilation for which the identity \eqref{eq-SFInd} can be checked by direct computation, and one which uses the index of Fredholm pairs of projections introduced by Kato \cite{Kat} and studied in detail in \cite{ASS}. 

\vspace{.2cm}

\noindent {\bf Proof of Theorem~\ref{theo-FredFlow} by homotopy.} Let $\Hh=\Kk\oplus\Kk'$ for some Hilbert space $\Kk'$. In this grading,
\begin{equation}
\label{eq:U&F}
U_T
\;=\;
\begin{pmatrix}
T & B \\ C & D
\end{pmatrix}
\;,\qquad\quad F
\;=\;
\begin{pmatrix}
\one & 0 \\ 0 & -\one
\end{pmatrix}
\;,
\end{equation}
with adequate operators $B$, $C$ and $D$. By unitarity, $BB^*=\one-TT^*$ and $C^*C=\one-T^*T$. As $T$ is essentially unitary, it follows that $BB^*$ and $C^*C$ are compact and thus also $B$ and $C$ are compact ({\it e.g.} by the polar decomposition). This implies that $[F,U_T]\in\KM(\Hh)$. Furthermore,  $F\in \widehat{\SM\FM}_*(\Hh)$ and thus by Proposition~\ref{prop-SFFU}  the spectral flow $\SF(F,U_T)$ is well-defined and the second equality in \eqref{eq-SFInd} holds. Because it was assumed that $\Kk'=\Hh\ominus\Kk$ is infinite dimensional, there exists a  unitary map from $\Kk$ to $\Kk'$ and, as a basis change does not change the spectral flow, one may suppose $\Kk=\Kk'$ from now on. Actually the infinite dimensionality of $\Kk'$ follows if $\Ind(T)\not=0$. Indeed, by unitarity of $U_T$ also $D^*D-\one=B^*B$ and $DD^*-\one=CC^*$ are compact, so that $D$ is also an essential unitary operator. Since the index of $U_T$ vanishes and the index is stable under compact perturbations, it follows that $0=\Ind(T\oplus D)=\Ind(T)+\Ind(D)$. But a non-vanishing index only exists in infinite dimension.

\vspace{.2cm}

The basic idea of the proof is to verify \eqref{eq-SFInd} for one particular unitary dilation $U_T^{\mbox{\rm\tiny H}}$ and then to show that any other unitary dilation $U_T$ is  connected with $U_T^{\mbox{\rm\tiny H}}$ by a continuous path $r\in[0,1]\mapsto U(r)$ which is such that $(F,U(r))\in\PM(\Hh)$ holds (which is equivalent to the off-diagonal entry of $U(r)$ being compact). By Proposition~\ref{prop-SFFU} one then has $\SF(F,U_T)=\SF(F,U_T^{\mbox{\rm\tiny H}})$ and one can conclude that \eqref{eq-SFInd} holds for any unitary dilation if only it can be checked for the special dilation $U_T^{\mbox{\rm\tiny H}}$. The latter is chosen to be the Halmos dilation \cite{Hal}:
\begin{equation}
\label{eq-Halmos}
U_T^{\mbox{\rm\tiny H}}
\;=\;
\begin{pmatrix}
T &  (\one-TT^*)^{\frac{1}{2}} \\ (\one-T^*T)^{\frac{1}{2}} & -T^*
\end{pmatrix}
\;.
\end{equation}
Let $T=V|T|$ be the polar decomposition with partial isometry $V$ such that $\Ker(V)=\Ker(T)$. Due to the continuous path $r\in [0,1]\mapsto V|T|^r$ and the stability of the index one has $\Ind(T)=\Ind(V)$. Let us consider the path  $r\in [0,1]\mapsto U(r) $ with
$$
U(r)
\;=\;
\begin{pmatrix}
V|T|^r &  (\one-V|T|^{2r}V^*)^{\frac{1}{2}} \\ (\one-|T|^rV^*V|T|^r)^{\frac{1}{2}} & -|T|^rV^*
\end{pmatrix}
\;.
$$
Indeed, $(F,U(r))\in\PM(\Hh)$ and the path connects $U(1)=U_T^{\mbox{\rm\tiny H}}$ to 
$$
U(0)
\;=\;
\begin{pmatrix}
V &  (\one-VV^*)^{\frac{1}{2}} \\ (\one-V^*V)^{\frac{1}{2}} & -V^*
\end{pmatrix}
\;.
$$
Now
$$
F+s\,U(0)^*[F,U(0)]
\;=\;
\begin{pmatrix}
\one & 0 \\ 0 & -\one
\end{pmatrix}
\,+\,
2s\,\begin{pmatrix}
-(\one-V^*V) & 0 \\ 0 & \one-VV^*
\end{pmatrix}
\;
$$
where one uses that $\one-V^*V$ and $\one-VV^*$ are the orthogonal projections on $\Ker(V)=\Ker(T)$ and $\Ker(V^*)=\Ker(T^*)$, respectively. These relations also imply that the spectral flow of $s\in[0,1]\mapsto F+s\,U(0)^*[F,U(0)]$   is equal to $-\Ind(T)$ and therefore the identity \eqref{eq-SFInd} for the Halmos dilation follows. 

\vspace{.2cm}

It now remains to construct a continuous map from an arbitrary dilation $U_T$ to $U_T^{\mbox{\rm\tiny H}}$ such that, when combined with $F$, one stays in $\PM(\Hh)$. For that purpose, let us consider $W=(U_T^{\mbox{\rm\tiny H}})^*U_T$ and show that it is homotopic to the identity with a path having compact off-diagonal. The matrix entries of $W =\begin{pmatrix} A & B \\ C & D \end{pmatrix}$ satisfy $\Ind(A)=\Ind(D)=0$ (actually, one also has $A=T^*T$, but this is irrelevant in the following) and $B$ and $C$ are compact. Consequently, by standard Fredholm theory there are partial isometries $V_A$ and $V_D$ such that $A+\epsilon\, V_A$ and $D+\epsilon \,V_D$ are invertible for $\epsilon>0$. Now choose $\epsilon\in(0,1)$ sufficiently small such that the path $r\in [0,\epsilon]\mapsto W(r) =\begin{pmatrix} A+r\,V_A & B \\ C & D +r\, V_D\end{pmatrix}$ remains in the invertibles (the spectrum of $W(r)$ does not touch $0$ if $r<1$). Next decompose
$$
W(r)
\;=\;
\begin{pmatrix} A+r\,V_A & 0 \\ 0 & D +r\, V_D\end{pmatrix}
\,
\begin{pmatrix} \one &  (A+r\,V_A)^{-1}B \\ (D +r\, V_D)^{-1}C & \one \end{pmatrix}
\;.
$$
The second factor is a compact perturbation of the identity and can therefore be continuously deformed within the compact operators to the identity, {\it e.g.} using spectral theory for compact operators. Thus one obtains a path $r\in[\epsilon,1]\mapsto W(r)$ in the invertibles on $\Kk\oplus\Kk$ with compact off-diagonal entries connecting $W(\epsilon)$ to the identity (as also the invertibles on $\Kk$ are path connected by the polar decomposition). From this path of invertibles one obtains the desired path of unitaries $r\in [0,1]\mapsto W(r)|W(r)|^{-1}$. Let us note that the homotopy is not a path of unitary dilations of $T$, but this is not needed in order to connect the spectral flows.
\hfill $\Box$

\vspace{.2cm}

\noindent {\bf Proof of Theorem~\ref{theo-FredFlow} using Fredholm pairs of projections.} (This proof looks shorter than the above, but appeals to several results from \cite{ASS}.) Let $U_T$ and $F$ be as in \eqref{eq:U&F} and consider any path $[0,1]\in s\to F_s$ in $\Theta(F,U_T)$, namely such that $F_0=F$, $F_1=U_T^*FU_T$ and $F_s-F\in\KM(\Hh)$ for all $s\in[0,1]$. Let us start from the definition \eqref{eq-SFdef} of the spectral flow. As $Q_{a_n,0}(s_{n-1})$ and $Q_{a_n,0}(s_n)$ are both finite dimensional projections, the difference of their traces can be expressed in terms of the index of a pair of projections. One of the equivalent definitions in \cite{ASS} is $\mbox{\rm Ind}(P,Q)=\mbox{\rm Ind}(QPQ)$ whenever $QPQ$ is a Fredholm operator on $Q\Hh$. With this, 
$$
\SF(s\in[0,1]\mapsto F_s)
\;=\;
\sum_{n=1}^N \;\mbox{\rm Ind}\left(Q_{a_n,0}(s_{n-1}), Q_{a_n,0}(s_n)\right)
\;.
$$
Let us consider the  orthogonal decomposition $P(s)=Q_a(s)\oplus Q_{a,0}(s)$ for $a<0$. The map $[s_{n-1},s_n]\ni s\mapsto Q_{a_n}(s)$ is continuous and for any pair $s_{n-1}\leqslant s_1'< s_2'\leqslant s_n$ an application of the Riesz integral and of the resolvent identity shows that the difference $Q_{a_n}(s_1')-Q_{a_n}(s_2')$ is compact and $\|Q_{a_n}(s_1')-Q_{a_n}(s_2')\|\leqslant C\ \|F_{s_1'}-F_{s_2'}\|$. From these two facts together follows that $Q_{a_n}(s_{n-1})$ and $Q_{a_n}(s_n)$ are a Fredholm pair with ${\rm Ind}(Q_{a_n}(s_{n-1}),Q_{a_n}(s_n))=0$ \cite[Proposition 3.2 and Theorem 3.4(c)]{ASS}, see also \cite[Lemma~4.1]{BCP} where a characterization for being a Fredholm pair is given. Since the computation of the index is linear with respect to the orthogonal sum of projections \cite[Lemma A.9]{ASS}, one has ${\rm Ind}(Q_{a_n,0}(s_{n-1}), Q_{a_n,0}(s_n))\;=\;{\rm Ind}(P(s_{n-1}), P(s_{n}))$. Observing that the difference $P(s_{n-1})- P(s_{n})$ is compact, one uses again \cite[Theorem 3.4(c)]{ASS} for the sum of the telescopic series:
$$
\SF(s\in[0,1]\mapsto F_s)\;=\; {\rm Ind}(P(0), P(1))\;=\; {\rm Ind}(P(0),U_T^* P(0)U_T)\;=\; {\rm Ind}(P(0)U_T P(0))\;
$$
where the last step is provided by \cite[Theorem 5.2]{ASS} and $P(0)U_T P(0)$ has to be considered as an operator on $P(0)\Hh$. Observing that $P(0)\oplus P=\one$, with $P=\chi_{[0,+\infty)}(F)$ the spectral projection of $F$ on the positive spectrum, one concludes that  ${\rm Ind}(P(0)U_T P(0))=-{\rm Ind}(PU_T P)=-{\rm Ind}(T)$ since  $PU_T P|_{P\Hh}=\Pi^*U_T\Pi=T$.
\hfill $\Box$

\section{Mapping cone of a Fredholm module}
\label{sec-mappingcone}

Let $\Aa$ be a C$^*$-algebra which is realized as a subalgebra of the bounded operators $\BM(\Hh)$ on a separable Hilbert space $\Hh$. Associated are the $K$-groups $K_0(\Aa)$ and $K_1(\Aa)$ of homotopy classes of projections and unitaries in $\Aa$. Topological content can be extracted from these groups via pairing with Fredholm modules. The ungraded version of an even bounded Fredholm module $(\Hh,F)$ for $\Aa$ is a unitary operator $F$ on $\Hh$ such that for all $A\in\Aa$ the commutators $[F,A]\in\KM(\Hh)$. An odd bounded Fredholm module is an even bounded Fredholm module $(\Hh,F)$ for which, moreover, $F^2=\one$ and $\sigma_\ess(F)=\{-1,1\}$. Equivalently, the unitary $F$ lies in $\widehat{\SM\FM}_*(\Hh)$. In the literature \cite{Con}, even Fredholm modules are always graded, so let us briefly explain the connection as well as the relation between unitary dilations and odd Fredholm modules.

\vspace{.2cm}

\noindent {\bf Remark 1} 
By definition \cite{Con}, a (graded version of an) even Fredholm module $(\widehat{\Hh},\widehat{F},\Gamma)$  for a C$^*$-algebra $\widehat{\Aa}$ is an odd Fredholm module $(\widehat{\Hh},\widehat{F})$ together with a (unitary) grading operator $\Gamma$ satisfying $\Gamma^2=\one$, $\Gamma A\Gamma=A$ for all $A\in\widehat{\Aa}$, and $\Gamma \widehat{F}\Gamma=-\widehat{F}$. Given an ungraded version of an even bounded Fredholm module $(\Hh,F)$ for $\Aa$,  one obtains a graded version by setting $\widehat{\Hh}=\Hh\otimes\CM^2$ and $\widehat{\Aa}=\Aa\otimes\one_2$, furthermore setting $\widehat{F}=\Re e(F)\,\sigma^1+\Im m(F)\,\sigma^2$ and $\Gamma=\sigma^3$ where $\sigma^1$, $\sigma^2$ and $\sigma^3$ are the Pauli matrices and $\Re e(F)=\frac{1}{2}(F+F^*)$ as well as $\Im m(F)=\frac{1}{2\imath}(F-F^*)$. Inversely, given a graded version of an even Fredholm module, the reduction to the spectral subspaces of $\Gamma$ allows to extract two ungraded versions of an even Fredholm module.
\hfill $\diamond$

\vspace{.2cm}

\noindent {\bf Remark 2} The operator $F$ of an odd Fredholm module is also called the Dirac phase because it can often be obtained from a so-called unbounded Fredholm module $(\Hh,D)$ with Dirac operator $D$. Here $D$ is an invertible, self-adjoint operator with compact resolvent and a domain $\Dd(D)$ which is left invariant under $\Aa$, and such that the commutators $[A,D]$ are compact operators for all $A\in\Aa$.  From this data one constructs a bounded Fredholm module by setting $F=D|D|^{-1}$, see \cite{Con,CP}. Similarly, the operator $\widehat{F}$ of a graded version of an even Fredholm module is often obtained from a Dirac operator.
\hfill $\diamond$

\vspace{.2cm}

\noindent {\bf Remark 3} Let $T\in\EM\UM(\Kk)$ be an essentially unitary Fredholm operator on a Hilbert space $\Kk$ and $U_T\in\UM(\Hh)$ a unitary dilation on a Hilbert space $\Hh$ in which there is an isometric embedding $\Pi:\Kk\mapsto\Hh$, see \eqref{eq-dildef}. Then setting $\Aa=C^*(U_T)$ and $F=2\,\Pi\Pi^*-\one$, one has a odd Fredholm module $(\Hh,F)$ for $\Aa$. Indeed, $F\in\widehat{\SM\FM}_*(\Hh)$ and the off-diagonal entries of $U_T$ are compact by the argument of the proof of Theorem~\ref{theo-FredFlow} which assures the required compactness. Inversely, if $(\Hh,F)$ is an odd Fredholm modul for $\Aa=C^*(U)$ for some unitary $U\in\BM(\Hh)$, then $U$ is a unitary dilation of $PUP$ where $P=\frac{1}{2}(F+\one)$.
\hfill $\diamond$

\vspace{.2cm}

It is well-known \cite[Chapter~4]{Con} that odd Fredholm modules pair integrally with the $K$-group $K_1(\Aa)$ via
\begin{equation}\label{eq:odd_pair}
\langle (\Hh,F)\,,\,[U]_1\rangle_1
\;=\;
\Ind(P\,UP)\;,
\qquad
P=\tfrac{1}{2}(F+\one)\;,
\end{equation}
where here $U\in\Aa$ and $P\,UP$ is understood as an operator on $\Kk=P\Hh={\rm Ran}(P)$, and a natural generalization holds if $U$ only lies in a matrix algebra over $\Aa$. This makes sense because $P\,UP$ is indeed a Fredholm operator (with pseudo-inverse $P\,U^*P$).  Similarly, ungraded versions of even Fredholm modules pair integrally with $K_0(\Aa)$. Explicitly, again only for a projection $P\in\Aa$,
\begin{equation}\label{eq:even_pair}
\langle (\Hh,F)\,,\,[P]_0\rangle_0
\;=\;
\Ind(PFP)\;,
\end{equation}
where $PFP$ is a Fredholm operator again on $\Kk=P\Hh=\Ran(P)$. The presentation already makes it clear that the two pairings are quite similar and connected via
$$
\langle (\Hh,F)\,,\,[P]_0\rangle_0
\;=\;
\langle (\Hh,2P-\one)\,,\,[F]_1\rangle_1
\;,
$$
where on the r.h.s. one has a pairing over the algebra $C^*(F)$.

\vspace{.2cm}

\noindent {\bf Remark 4} If one works with the graded version of an even Fredholm module $(\widehat{\Hh},\widehat{F},\Gamma)$  for a C$^*$-algebra $\widehat{\Aa}$, the even pairing takes the familiar form \cite{Con}
$$
\langle (\widehat{\Hh},\widehat{F},\Gamma)\,,\,[\widehat{P}]_0\rangle_0
\;=\;
\Ind(P_-FP_+)\;,
\qquad
\widehat{F}=
\begin{pmatrix}
0 & F^* \\ F & 0
\end{pmatrix}
\;,
\;\;\;
\widehat{P}=
\begin{pmatrix}
P_+ & 0 \\ 0 & P_-
\end{pmatrix}
\;,
$$
where $\widehat{F}$ and $\widehat{P}$ are written in the grading of $\Gamma$ and $P_-FP_+$ is understood as operator from ${\rm Ran}(P_+)$ to ${\rm Ran}(P_-)$. If $P_+=P_-$ this reduces the formula above  \eqref{eq:even_pair}.
\hfill $\diamond$

\vspace{.2cm}

\vspace{.2cm}

It is possible to calculate these two indices from the Chern characters constructed from the corresponding Fredholm modules  \cite{Con}, but here the aim will rather be to link these indices to adequate spectral flows. For this purpose, let us define the $F$-twisted mapping cone associated to $(\Hh,F)$ by
$$
\Mm
\;=\;
\left\{
(A_s)_{s\in[0,1]}\in C([0,1],\Aa+\KM(\Hh))\,\left|\,A_0=F^*A_1F\in\Aa\;\;\mbox{\rm and }\;K_s\;=\;A_s-A_0\in\KM(\Hh)\;
\right.\right\}
\;.
$$
This is a C$^*$-algebra and the evaluation application $\ev:\Mm\to\Aa$ defined by $\ev((A_s)_{s\in[0,1]})=A_0$ is an algebra homomorphism. Furthermore the suspension $S\KM(\Hh)=C_0((0,1),\KM(\Hh))$ is a subalgebra embedded in $\Mm$ so that $S\KM(\Hh)\;\hookrightarrow\;\Mm\;\rightarrow\;\Aa$ is exact. Finally, an explicit lift of $A\in\Aa$ into $\Mm$ is given by $(1-s)A+s\,F^*AF=A+s\,F^*[A,F]$. An arbitrary lift of $A\in\Aa$ is denoted by $\mbox{\rm Lift}(A)=(\mbox{\rm Lift}(A)_s)_{s\in[0,1]}$. If $A$ is selfadjoint, one can always assume the lift to be selfajoint as well (just work with $\frac{1}{2}(\mbox{\rm Lift}(A)_s^*+\mbox{\rm Lift}(A)_s)$ if necessary). Resuming, one has:

\begin{proposi}
\label{prop-exactsequ} The sequence $0\,\rightarrow\,S\KM(\Hh)\,\hookrightarrow\,\Mm\,\stackrel{\mbox{\rm\tiny ev}\;}\rightarrow\,\Aa\,\rightarrow\,0$ is exact.
\end{proposi}

\vspace{.2cm}

This sequence is splitting if and only if $F$ commutes with $\Aa$. The $6$-term exact sequence of $K$-theory associated to the exact sequence is
%
%
$$
\begin{aligned}
& K_0(S\KM(\Hh)) & 
\overset{\imath_\ast}{\longrightarrow} 
& & K_0(\Mm)) & & \overset{{\rm ev}_\ast}{\longrightarrow}& & K_0(\Aa)&
\\
\\
& \Ind\;\;\uparrow  & & & & & & & \downarrow\;\;\exp&
\\
\\
& K_1(\Aa)  & \stackrel{\;\;{\mbox{\rm\tiny ev}_*}}{\longleftarrow} & & K_1(\Mm) & & \overset{\imath_\ast}{\longleftarrow} & & K_1(S\KM(\Hh))&
\end{aligned}
$$
Here $K_0(S\KM(\Hh))=K_1(\KM(\Hh))=0$ and $K_1(S\KM(\Hh))=K_0(\KM(\Hh))=\ZM$.  The connecting map $\exp:K_0(\Aa) \to K_1(S\KM(\Hh))$ on the r.h.s. is, in terms of a (self-adjoint) lift, explicitly given by
\begin{equation}
\label{eq-expmap}
\exp([P]_0)
\;=\;
[\exp(2\pi\imath \;\mbox{\rm Lift}(P)_{{s}})]_1
\;=\;
[\exp(2\pi\imath (P+s\,F^*[P,F]))]_1
\;,
\end{equation}
where in the second equality an explicit lift was chosen. Furthermore, the index map from $K_1(\Aa)$ to $K_0(S\KM(\Hh))=0$ is trivial. Thus the other map of interest to us  $\ev_*:K_1(\Mm)\to K_1(\Aa)$ is surjective and has an inverse $\ev^*:K_1(\Aa)\to K_1(\Mm)$. Therefore, every unitary $U\in\Aa$ has a unitary lift $\mbox{\rm Lift}(U)\in\Mm$. In general, it may be difficult to find an explicit formula for such a unitary lift, but if the Fredholm module is odd then an explicit choice is given by
\begin{align}
\ev^*([U]_1)
& = \;
\bigl[U\,\exp\bigl(\tfrac{\imath\pi}{2}\bigl(-\one+(1-s)F+s\,U^*FU\bigr)\bigr)\,F\bigr]_1
\nonumber
\\
& =\;
\bigl[U\,\exp\bigl(\tfrac{\imath\pi}{2}\bigl(F-\one+s\,U^*[F,U]\bigr)\bigr)\,F\bigr]_1
\;.
\label{eq-liftchoice}
\end{align}
A further difference between odd and ungraded even Fredholm modules is pointed out in the following result.

\begin{proposi}
\label{prop-KtheoryMappingCone} If $(\Hh,F)$ is odd, then $\exp=0$, namely every class of  projections in $\Aa$ is the image of a class of  projections in $\Mm$. 
\end{proposi}

\noindent {\bf Proof.} If the Fredholm module is odd, then it is possible to choose a lift given by a path of projections. By \eqref{eq-expmap} this implies the result. Let $Q=\frac{1}{2}(F+\one)$ be the positive spectral projection of $F$. Then set
$$
\mbox{\rm Lift}(P)_s
\;=\;
QPQ\,+\,
e^{\imath \pi s}\,QP(\one-Q)\,+\,
e^{-\imath \pi s}\,(\one-Q)PQ+(\one-Q)P(\one-Q)
\;,
$$
which can indeed be checked to be self-adjoint and idempotent.
\hfill $\Box$

\vspace{.2cm}

Resuming, for an odd Fredholm module and a unitary $U$ one has $\ev^*([U]_1)\in K_1(\Mm)$, while for an even Fredholm module and a projection $P$ one has $\exp([P]_0)\in K_1(S\KM(\Hh))$. In both cases, there is a spectral flow associated. Let us begin with the odd case. The path of unitaries $s\in[0,1]\mapsto V_s=U^*\mbox{\rm Lift}(U)_sF$ satisfies $V_0=F$ and $V_1=U^*FU$. Hence the spectrum of $V_0$ and $V_1$ consists of two infinitely degenerate eigenvalues $\{-1,1\}$ so that $V_0,V_1\in\widehat{\SM\FM}_*(\Hh)$. As $V_s$ is a compact perturbation of $V_0$, there are a finite number of eigenvalues (on the unit circle) moving from $-1$ to $1$ along the path $s\in[0,1]\mapsto V_s$ and this allows to define a spectral flow. In order to formulate this spectral flow again in the picture of paths of self-adjoint operators presented in Section~\ref{sec-specflow}, one can consider $\Re e(V_s)=\frac{1}{2}(V_s-V_s^*)$ the spectrum of which consists of the spectrum of $V_s$ projected onto the real line (along the imaginary axis). Thus
$$
\SF(s\in[0,1]\mapsto U^*{\rm Lift}(U)_sF)
\;=\;
\SF(s\in[0,1]\mapsto \Re e(V_s))
\;=\;
\SF(F,U)
\;,
$$
where the last equality reflects $\Re e(V_0)=F$ and $\Re e(V_1)=U^*FU$. Theorem~\ref{theo-FredFlow} combined with the above discussion then implies the following.

\begin{theo}
\label{theo-pairingsodd} Let $U\in\Aa$ and $(\Hh,F)$ be an odd Fredholm module.  Then
$$
\langle (\Hh,F)\,,\,[U]_1\rangle_1
\;=\;
\SF(F,U)
\;.
$$
\end{theo}

Next let us consider the even case. First of all,  $\exp(2\pi\imath\,{\rm Lift}(P)_s)-\one\in\KM(\Hh)$ so that the essential spectrum of $\exp(2\pi\imath\,{\rm Lift}(P)_s)$ is $\{1\}$ for all $s$ independently of the choice of  the self-adjoint lift, namely $\exp(2\pi\imath\,{\rm Lift}(P)_s)\in\UM_*(\Hh)$. Therefore the path has a winding number
$$
\mbox{\rm Wind}(s\in[0,1]\mapsto \exp(2\pi\imath\,{\rm Lift}(P)_s))
\;=\;
\SF(2P-\one,F)
\;.
$$

\begin{theo}
\label{theo-pairingseven} Let $P\in\Aa$ and $(\Hh,F)$ an even Fredholm module for $\Aa$. Then
$$
\langle (\Hh,F)\,,\,[P]_0\rangle_0
\;=\;
\SF(2P-\one,F)
\;.
$$
\end{theo}

\section{Real and quaternionic structures}
\label{sec-struct}

Let $\Hh$ be a complex seperable Hilbert space with complex conjugation $\Cc$ (also called real structure), namely $\Cc$ is anti-unitary and $\Cc^2=\one$. This allows to define the complex conjugate $\overline{T}=\Cc T\Cc\in\BM(\Hh)$ for any operator $T\in\BM(\Hh)$ as well as its transpose $T^t=(\overline{T})^*$. As a further structure on $\Hh$ be given unitary operators $J,I\in\BM(\Hh)$ which are real $J=\overline{J}$ and $I=\overline{I}$, and satisfy $J^2=\one$ as well as $I^2=-\one$. Note that $J=\one$ is a possible choice, and that the existence of $I$ forces a finite dimensional $\Hh$ to be of even dimension. Then $J\Cc$ and $I\Cc$ are respectively called real and quaternionic structures on $\Hh$. 

\begin{defini}
Let be given the above data and an operator $T\in\BM(\Hh)$.

\vspace{.1cm}

\noindent {\rm (i)} $T$ is even real {\rm(}or simply real{\rm )} $\Longleftrightarrow$ $J^*\overline{T}J=T$.

\vspace{.1cm}

\noindent {\rm (ii)} $T$ is odd real {\rm(}or quaternionic{\rm )}  $\Longleftrightarrow$ $I^*\overline{T}I=T$.

\vspace{.1cm}

\noindent {\rm (iii)} $T$ is even symmetric $\Longleftrightarrow$ $J^*T^tJ=T$.

\vspace{.1cm}

\noindent {\rm (iv)} $T$ is odd symmetric $\Longleftrightarrow$ $I^*T^tI=T$.

\vspace{.1cm}

\noindent The sets of even and odd real operators are denoted by $\BM\RM(\Hh,J)$ and $\BM\RM(\Hh,I)$, those of even and odd symmetric operators by $\BM\SM(\Hh,J)$ and $\BM\SM(\Hh,I)$. Similarly, the even and odd real Fredholm operators are $\FM\RM(\Hh,J)$ and $\FM\RM(\Hh,I)$ and the even and odd symmetric Fredholm operators $\FM\SM(\Hh,J)$ and $\FM\SM(\Hh,I)$.
\end{defini}

Via $T\in\BM\SM(\Hh,I)\mapsto IT$ the odd symmetric operators are in bijection with the anti-symmetric operators, and similarly the odd symmetric Fredholm operators are in bijection with the anti-symmetric Fredholm operators (which are not necessarily real as in \cite{AS} though). The reason to put forward the odd symmetric operators is that they have even degeneracies at every level of the Jordan hierarchy \cite{Sch}, a property that does not hold for antisymmetric operators. The following theorem can be deduced from \cite{AS}. A detailed, purely functional analytic proof is given in \cite{Sch}.

\begin{theo}
\label{theo-indices}
Let be given the above data. 

\vspace{.1cm}

\noindent {\rm (i)} The connected components of $\FM\RM(\Hh,J)$ are labelled by $T\in \FM\RM(\Hh,J)\mapsto\Ind(T)\in\ZM$ which 

is a  homotopy invariant with $\Ind(T+K)=\Ind(T)$ for compact $K\in\BM\RM(\Hh,J)$.

\vspace{.1cm}

\noindent {\rm (ii)} The connected components of $\FM\RM(\Hh,I)$ are labelled $T\in \FM\RM(\Hh,I)\mapsto\Ind(T)\in2\,\ZM$ which

is a  homotopy invariant with $\Ind(T+K)=\Ind(T)$ for compact $K\in\BM\RM(\Hh,I)$.

\vspace{.1cm}

\noindent {\rm (iii)} $\FM\SM(\Hh,J)$ has one connected component.

\vspace{.1cm}

\noindent {\rm (iv)} The $\ZM_2$ index $T\in\FM\SM(\Hh,I)\mapsto\Ind_2(T)=\dim(\Ker(T))\;\mbox{\rm mod}\,2\in\ZM_2$ is a homotopy invariant 

which labels the two connected components of $\FM\SM(\Hh,I)$ and satisfies $\Ind(T+K)=\Ind(T)$ 

for compact $K\in\BM\SM(\Hh,I)$.

\end{theo}

Items (i) and (ii) are relatively straight-forward, with (ii) following from a Kramers' degeneracy argument. More involved are the proofs of (iii) and (iv), with (iv) pending on the above mentioned even spectral degeneracy of odd symmetric matrices. Furthermore, the following representation theorem for odd symmetric operators of \cite{Sch} (generalizing a result of C.~L.~Siegel) is crucial for the proof of (iv) and will also be used below. Note that it also implies that the sets of odd symmetric invertibles and odd symmetric unitaries are path connected. 

\begin{theo}
\label{theo-Siegel}
$T\in\BM\SM(\Hh,I)$ $\Longleftrightarrow$ there exists $A\in\BM(\Hh)$ with $T=I^*A^tIA$.
\end{theo}

\noindent {\bf Examples} Let us provide examples of Fredholm operators in the above classes with non-trivial indices. Let $S$ be the left shift on the Hilbert space $\Hh=\ell^2(\NM)$ furnished with the natural fiberwise complex conjugation $\Cc$ and  $J=\one$. Then $S^n\in\FM\RM(\Hh,J)$ and $\Ind(S^n)=n$. Next let us consider $\Hh=\ell^2(\NM)\otimes\CM^2$ with $I=\binom{0\;\,-\one}{\one\;\;\;0}$. Then $T=\binom{S^n\;0}{0\;\;S^n}\in\FM\RM(\Hh,I)$ and $T'=\binom{S\;\;\;0}{0\;\;S^*}\in\FM\SM(\Hh,I)$ with $\Ind(T)=2n$ and $\Ind_2(T')=1$.
\hfill $\diamond$

\vspace{.2cm}

In view of Section~\ref{sec-specflowFred}, it is now natural to consider dilations of Fredholm operators in $\FM\RM(\Hh,J)$, $\FM\RM(\Hh,I)$, $\FM\SM(\Hh,J)$ and $\FM\SM(\Hh,I)$, which have the same symmetry properties, and then to study the associated spectral flow as in Theorem~\ref{theo-FredFlow}. For $\FM\RM(\Hh,J)$ and $\FM\RM(\Hh,I)$ there is little new to discover, except that the spectral flows associated to operators $\FM\RM(\Hh,I)$ all have Kramers degeneracy. More interesting is the case of odd symmetric Fredholm operators for which the concept of $\ZM_2$-valued spectral flow will be introduced in the next section. This is then connected to $\ZM_2$ indices in Section~\ref{sec-oddsympSF}.

\section{$\ZM_2$-valued spectral flow }
\label{sec-Z2flow}

The aim of this section is to introduce the new concept of a $\ZM_2$ valued spectral flow $\SF_2(F,U)$ similar as in \eqref{eq-SFFU}, but under the supplementary assumption that $F$ and $U$ are odd symmetric. Hence let $F\in\widehat{\SM\FM}_*(\Hh,I)=\widehat{\SM\FM}_*(\Hh)\cap\BM\SM(\Hh,I)$ and $U\in\UM\SM(\Hh,I)=\UM(\Hh)\cap\BM\SM(\Hh,I)$. Then let $\Theta(F,U,I)$ be the set of paths in $s\in[0,1]\mapsto F_s\in \widehat{\SM\FM}_*(\Hh)$ such that 
\begin{align}
{\rm (i) } \;& \;\;\;\; F_0=F\;,
\nonumber
\\
{\rm (ii) }\;& \;\;\;\; F_s-F\in \KM(\Hh) \;\;\mbox{\rm for all } s\in[0,1]\;,
\label{eq-SFcondOdd}
\\
{\rm (iii)' } & \;\;\;\; F_{1-s}\,=\,(IU)^*(F_s)^t(IU)\;.
\nonumber
\end{align}
Let us note that conditions (i) and (ii) are the same as in \eqref{eq-SFcond}, but (iii)' is strictly stronger because (iii)' indeed implies $F_1=U^*FU$. Furthermore, it implies $\sigma(F_{1-s})=\sigma(F_s)$. Hence the spectral curves are already determined by the first half $s\in[0,\frac{1}{2}]\mapsto F_s$ of the path. A particular path realizing (iii)' is given by $s\in[0,1]\mapsto F+s\,U^*[F,U]$, as can readily be checked. The odd symmetry assumptions imply Kramers' degeneracy of $F_s$ for $s=0,\frac{1}{2},1$ by the following result:

\begin{lemma}
\label{lem-Kramers}  
$F_0$, $F_{\frac{1}{2}}$ and $F_1$ have even degeneracy for every finitely degenerate eigenvalue.
\end{lemma}

\noindent {\bf Proof.} The arguments being similar, let us focus on the case $s=\frac{1}{2}$. One has $F_{\frac{1}{2}}V=V\overline{F}_{\frac{1}{2}}$ where $V=U^*I^*$. If $F_{\frac{1}{2}}\psi=\lambda \psi$ for some $\psi\in\Hh$ and $\lambda\in\RM$, then $F_{\frac{1}{2}}(V\overline{\psi})=V\overline{(F_{\frac{1}{2}}\psi)}={\lambda}\; (V\overline{\psi})$.
It remains to show that $V\overline{\psi}$ and $\psi$ are linearly independent. Suppose that there is some $a\in\CM$ such that $a\psi=V\overline{\psi}$. Then $|a|^2\psi=\overline{a}\,V\overline{\psi}=V\overline{V}\psi=-\psi$ which implies $a=0$. This shows that $\lambda$ is at least twice degenerate. If the degeneracy is larger, then one chooses an eigenvector in the orthogonal complement of the span of $\psi$ and $V\overline{\psi}$ and runs the argument above. Repeating this procedure proves the lemma.
\hfill $\Box$

\vspace{.2cm}

\begin{proposi}
\label{prop-SFFUodd} 
Let $F\in\widehat{\SM\FM}_*(\Hh,I)$ and $U\in\UM\SM(\Hh,I)$. Then  
$$
\SF(s\in[0,\tfrac{1}{2}]\mapsto F_s)
\;\;\mbox{\rm mod}\;2
\;,
$$
is independent of the choice of path $s\in[0,1]\mapsto F_s$ in $\Theta(F,U,I)$ and thus defines $\SF_2(F,U)\in\ZM_2$ which is called a $\ZM_2$ spectral flow. In particular, it can be calculated by a special path:
\begin{equation}
\label{eq-SF2def}
\SF_2(F,U)
\;=\;
\SF(s\in[0,\tfrac{1}{2}]\mapsto F+s\,U^*[F,U])
\;\;\mbox{\rm mod}\;2
\;.
\end{equation}
\end{proposi}

\noindent {\bf Proof.} Note that the definition of $\SF_2(F,U)$ only invokes half of the spectral curves, reflecting that $\sigma(F_s)=\sigma(F_{1-s})$. Let $s\in[0,1]\mapsto F_s$ and $s\in[0,1]\mapsto F'_s$ be two paths in $\Theta(F,U,I)$. Then $r\in[0,1]\mapsto rF_s+(1-r)F'_s$ is a homotopy in $\Theta(F,U,I)$ keeping the initial and final point fixed. It has to be shown that $\SF(s\in[0,\tfrac{1}{2}]\mapsto F_s)\,\mbox{\rm mod}\,2$ is constant during this homotopy. The spectral flow $\SF(s\in[0,\tfrac{1}{2}]\mapsto F_s)$ is not necessarily constant because the right end point $F_{\frac{1}{2}}$ is not being kept fixed (even though the left end point $F_0$ is fixed), but it only changes by multiples of $2$ because of the Kramers' degeneracy of $F_{\frac{1}{2}}$ proved in Lemma~\ref{lem-Kramers}.
\hfill $\Box$

\vspace{.2cm}

\noindent {\bf Remark} 
If $0\not\in\sigma(F_s)$ for $s=0,\frac{1}{2}$ and the spectral curves of $s\in[0,\frac{1}{2}]\mapsto F_s$ intersect the segment $[0,1]\times\{0\}$ transversally, $\SF_2(F,U)$ can simply be calculated as the number of intersections with the segment modulo $2$. Note that $\SF(F,U)=0$.
\hfill $\diamond$

\vspace{.2cm}

The following proposition states a consequence of non-trivial $\ZM_2$-spectral flows. It is readily proved by contraposition. 

\begin{proposi}
\label{prop-SFFUoddnontrivial} 
Let $F\in\widehat{\SM\FM}_*(\Hh,I)$ and $U\in\UM\SM(\Hh,I)$. Suppose that $\SF_2(F,U)=1$ and let $s\in[0,1]\mapsto F_s$ be any path satisfying {\rm \eqref{eq-SFcondOdd}}. Then there exists at least one evenly degenerate eigenvalue $\lambda\in\sigma(F_{\frac{1}{2}})\cap (-1,1)$.
\end{proposi}

Let us introduce the set of operator pairs for which a $\ZM_2$ spectral flow can be defined:
$$
\PM(\Hh,I)
\;=\;
\left\{\left.
(F,U)\in\widehat{\SM\FM}_*(\Hh,I)\times \UM(\Hh,I)
\;\right|\;
[F,U]\;\mbox{\rm compact}
\right\}
\;.
$$
It carries the subspace topology induced from the norm topology on $\BM(\Hh)\times\BM(\Hh)$. As $\SF(F,U)=0$ for all $(F,U)\in\PM(\Hh,I)$, one has $ \PM(\Hh,I)\subset\PM(\Hh)_0$. Note that indeed each point $(F,U)\in\PM(\Hh,I)$ defines a class of paths $\Theta(F,U,I)$ so that it is possible to view the $\ZM_2$ spectral flow as a map $\SF_2:\PM(\Hh,I)\to \ZM_2$.

\begin{proposi}
\label{prop-SF2FU2} 
The spectral flow $\SF_2:\PM(\Hh,I)\to \ZM_2$ is locally constant. 
\end{proposi}

\noindent {\bf Proof.} Let $r\in[0,1]\mapsto (F(r),U(r))\in\PM(\Hh,I)$ be a homotopic change in $\PM(\Hh,I)$. Associated are the paths $s\in[0,\tfrac{1}{2}]\mapsto F(r)+s\,U(r)^*[F(r),U(r)]$ for each of which $\SF_2(F(r),U(r))$ is well-defined by Proposition~\ref{prop-SFFUodd}. Now the homotopic deformations change the spectrum of $F_0=F$ and $F_{\frac{1}{2}}$, but by Lemma~\ref{lem-Kramers} only by multiples of $2$ each. Thus $\SF_2(F(r),U(r))$ is constant in $r$.
\hfill $\Box$

\section{Odd symmetric dilations and their $\ZM_2$ spectral flow}
\label{sec-oddsympSF}

In agreement with the previously introduced notation let us denote by $\EM\UM\SM(\Hh,I)$ the subset of odd symmetric  essentially unitary Fredholm operators.

\begin{theo}
\label{theo-OddFredFlow} Let $T\in\EM\UM\SM(\Kk,I_\Kk)$ be an odd symmetric essentially unitary operator on a separable Hilbert space $\Kk$ furnished with an orthogonal $I_\Kk$ satisfying $(I_\Kk)^2=-\one$. Suppose that $\|T\|\leq 1$. Let $U_T\in\BM\SM(\Hh,I_\Hh)$ be an odd symmetric unitary dilation with associated injective partial isometry $\Pi:\Kk\to\Hh$ satisfying $I_\Kk=\Pi^* I_\Hh \Pi$. Here $I_\Hh$ is an orthogonal on $\Hh$ with $(I_\Hh)^2=-\one$. Then, with $F=2\,\Pi\Pi^*-\one$,
\begin{equation}
\label{eq-oddSFInd}
\Ind_2(T)
\; =\;
\SF_2\bigl(F,U_T \bigr)
\; =\;
\SF_2\bigl(s\in[0,1]\mapsto F+s\,(U_T)^*[F,U_T] \bigr)
\;.
\end{equation}
\end{theo}

\noindent {\bf Proof.} Let $\Hh=\Kk\oplus\Kk'$ furnished with a real unitary $I_\Hh=I_\Kk\oplus I_{\Kk'}$ satisfying $(I_\Hh)^2=-\one$. Both $\Kk$ and $\Kk'$ are infinite dimensional (and separable) and therefore Proposition~5 of \cite{Sch} implies that there exists an orthogonal (namely a real unitary) $O:\Kk\to\Kk'$ such that $I_{\Kk'}=OI_\Kk O^t$. Hence conjugating $U_T$ with $\mbox{\rm diag}(\one,O)$ leads to a new odd symmetric dilation still denoted by $U_T$ which now acts on $\Hh=\Kk\oplus\Kk$ furnished with $I_\Hh=I_\Kk\oplus I_\Kk$. From now on the index on $I$ will be suppressed. In the grading of $\Hh=\Kk\oplus\Kk$ the dilation has to have the form
$$
U_T
\;=\;
\begin{pmatrix}
T & B \\ I^*B^tI & D
\end{pmatrix}
\;.
$$
The unitarity of $U_T$ is equivalent to the $4$ equations
$$
TT^*+BB^*\;=\;\one_{\Kk}\;,
\quad
B^*B+D^*D\;=\;\one_{\Kk}\;,
\quad
T^*B+I^*\overline{B}ID\;=\;0\;,
\quad
TI^*\overline{B}I+BD^*\;=\;0\;.
$$
This shows again that $B$ is compact. As $\Ind_2(U_T)=0$, it thus follows by homotopy that $\Ind_2(D)=\Ind_2(T)$. Let us note that the Halmos dilation $U^{\mbox{\rm\tiny H}}_T$ defined in \eqref{eq-Halmos} lies in $\UM\SM(\Hh,I)$. As a first step, it will be verified that \eqref{eq-oddSFInd} holds for $U^{\mbox{\rm\tiny H}}_T$. Two cases are considered separately. First of all, if $\Ind_2(T)=0$, then by Theorem~\ref{theo-indices}(iv) there exists a path in $\FM\SM(\Kk,I)$ from $T$ to the identity. This leads to a path of Halmos dilations (and thus, in particular, in $\UM\SM(\Hh,I)$) connecting $U^{\mbox{\rm\tiny H}}_T$ to $\one$. Because the $\ZM_2$ spectral flow is a homotopy invariant, this implies in this case
$$
\SF_2\bigl(s\in[0,1]\mapsto F+s\,(U^{\mbox{\rm\tiny H}}_T)^*[F,U^{\mbox{\rm\tiny H}}_T] \bigr)
\;=\;
0\;=\;\Ind_2(T)
\;.
$$
Second of all, if $\Ind_2(T)=1$, again by Theorem~\ref{theo-indices}(iv), there exists a homotopy from $T$ to
$$
T_0\;=\;
\begin{pmatrix}
S & 0 \\ 0 & S^*
\end{pmatrix}
\;,
\qquad
I\;=\;
\begin{pmatrix}
0 & -\one \\ \one & 0
\end{pmatrix}
\;,
$$
where $S$ is a unilateral left shift on $\Kk$ of multiplicity $1$ (associated with an orthonormal sequence in $\Kk$ spanning half of $\Kk$). It satisfies $S^*S=\one-P$, $SS^*=\one$ and $SP=0$ where $P$ is a one-dimensional orthogonal projection on $\Kk$. Thus there exists a path of odd symmetric unitary dilations from $U^{\mbox{\rm\tiny H}}_T$ to 
$$
U_0
\;=\;
\begin{pmatrix}
S & 0 & 0 & 0 \\
0 & S^* & 0 & P \\
P & 0 & -S^* & 0 \\
0 & 0 & 0 & -S
\end{pmatrix}
\;.
$$
%
For this dilation, one has
$$
F+s\,(U_0)^*[F,U_0]
\;=\;
\begin{pmatrix}
\one & 0 & 0 & 0 \\
0 & \one & 0 & 0 \\
0 & 0 & -\one & 0 \\
0 & 0 & 0 & -\one
\end{pmatrix}
\;+\;
2\,s\,
\begin{pmatrix}
P & 0 & 0 & 0 \\
0 & 0 & 0 & 0 \\
0 & 0 & 0 & 0 \\
0 & 0 & 0 & -P
\end{pmatrix}
\;.
$$
Now one can read off
$$
\SF_2\bigl(s\in[0,1]\mapsto F+s\,(U_0)^*[F,U_0]\;\;\mbox{\rm by }0 \bigr)
\;=\;
1\;=\;\Ind_2(T)
\;.
$$
Invoking again the homotopy invariance of the $\ZM_2$ spectral flow and combining with the case $\Ind_2(T)=0$, one concludes that \eqref{eq-oddSFInd} holds for $U^{\mbox{\rm\tiny H}}_T$.

\vspace{.2cm}

Similar as in the first proof of Theorem~\ref{theo-FredFlow}, it now remains to construct a homotopy from $(F,U_T)$ to $(F,U_T^{\mbox{\rm\tiny H}})$ in $\PM(\Hh,I)$. Let us begin by factorizing $U_T$. From $TT^*+BB^*=\one$, it follows that there is a unitary $V$ such that
$$
B^*\;=\;V(\one-TT^*)^{\frac{1}{2}}
\;.
$$
This is {\it not} the polar decomposition in which $V$ would be a partial isometry with $\Ker(V)=\Ker(B^*)$, but rather a modification of it (which is arbitrary). Then
\begin{align}
U_T
& =\;
\begin{pmatrix}
T & (\one-TT^*)^{\frac{1}{2}}V^*
\\
I^*\overline{V}I\,I^*((\one-TT^*)^{\frac{1}{2}})^tI & D
\end{pmatrix}
\nonumber
\\
& = \;
I^*
\begin{pmatrix}
I^* & 0 \\
0 & I^*V^*
\end{pmatrix}^t
\begin{pmatrix}
T & (\one-TT^*)^{\frac{1}{2}}
\\
(\one-T^*T)^{\frac{1}{2}} & I^*V^tIDV
\end{pmatrix}
I
\begin{pmatrix}
I^* & 0 \\
0 & I^*V^*
\end{pmatrix}
\;.
\label{eq-factorizeUT}
\end{align}
Now recall that if $T$ is odd symmetric, then also $I^*A^tTIA$ for an arbitrary operator $A$. This shows, first of all, that $D'=I^*V^tIDV$ is odd symmetric. As all factors in \eqref{eq-factorizeUT} are unitary, also the matrix in the middle is an odd symmetric unitary dilation of $T$. Second of all, the factorization \eqref{eq-factorizeUT} shows explicitly that $U_T$ is odd symmetric (which, of course, was already known). This remains true if $V$ in the outer factors of \eqref{eq-factorizeUT} (not in the factor in the middle) is homotopically deformed to $\one$. Hence $U_T$ is homotopic to the odd symmetric unitary dilation
$$
U'_T
\;=\;
\begin{pmatrix}
T & (\one-TT^*)^{\frac{1}{2}}
\\
(\one-T^*T)^{\frac{1}{2}}  & D'
\end{pmatrix}
\;,
$$
and, moreover, during the homotopy the off-diagonal operators remain compact. For sake of simplicity, let us set $D'=D$. Now the unitarity of $U'_T$ implies
\begin{equation}
\label{eq-twoid}
TT^*\;=\;D^*D\;,
\qquad
(T^*+D)(\one-TT^*)^{\frac{1}{2}}\;=\;0
\;.
\end{equation}
Let us introduce the subspace $\Vv=\Ker(\one-TT^*)$ and denote by $P_\Vv$ the orthogonal projection on $\Vv$. Then $I^*(\one-T^*T)^tI=\one-TT^*$ shows $I\overline{\Vv}=\Ker(\one-T^*T)$. Note that also $\Vv=\Ker(\one-DD^*)$ and $I\overline{\Vv}=\Ker(\one-DD^*)$. Furthermore, $T^*:\Vv\mapsto I\overline{\Vv}$ because $v=TT^*v$ implies $T^*v=T^*T(T^*v)$. Similarly $T:I\overline{\Vv}\mapsto\Vv$, $D:\Vv\mapsto I\overline{\Vv}$ and $D^*:I\overline{\Vv}\mapsto\Vv$. Moreover, all four maps are isometric and therefore also unitary. Finally, $T^*:\Vv^\perp\to(I\overline{\Vv})^\perp$ as one checks by writing out the orthogonality relations.

\vspace{.1cm}

Using functional calculus of a unitary operator, let us now define unitaries $(T^*)^{\frac{1}{2}}:\Vv\to I\overline{\Vv} $ and $ D^{-\frac{1}{2}}:I\overline{\Vv}\to\Vv$ (here the isomorphism $I\overline{\Vv}\cong\Vv$ is suppressed in the notation). Thus $W|_\Vv=\imath\, D^{-\frac{1}{2}}(T^*)^{\frac{1}{2}}|_\Vv:\Vv\to\Vv$ is unitary. It is extended to a unitary on $\Kk$ by setting $W=W|_\Vv+(\one-P_\Vv)$. Let us choose a path $r\in[0,1]\mapsto W(r)|_\Vv$ of unitaries on $\Vv$ such that $W(0)|_\Vv=W|_\Vv$ and $W(1)|_\Vv=\one_\Vv$. Set $W(r)=W(r)|_\Vv+(\one-P_\Vv)$ and $D(r)=I^*W(r)^tIDW(r)$. First of all, one readily checks that $D(r)$ is odd symmetric for all $r$. Obviously $D(1)=D$. Moreover, with some care one also checks $D(0)|_\Vv=-T^*|_\Vv$ by construction. Finally $D(r)$ leaves $\Vv^\perp$ invariant and $D(r)|_{\Vv^\perp}=-T^*|_{\Vv^\perp}$ due to the second identity in \eqref{eq-twoid}. Hence $D(0)=-T^*$ so that $U'_T$ is homotopic to the Halmos projection inside $\UM\SM(\Hh,I)$.
\hfill $\Box$

\section{$\ZM_2$ mapping cone of a Fredholm module}
\label{sec-mappingconeodd}

Let $\Aa$ be a C$^*$-algebra of bounded operators on a complex Hilbert space $\Hh$ and suppose given an involutive *-antiautomorphism $\tau:\Aa\to\Aa$. Then $\Aa_\tau=\{A\in\Aa\,|\,\tau(A)=A\}$ is a real C$^*$-algebra which has 8 Real $K$-groups \cite{Sc}. Here the focus will be on $\tau$ of the form $\tau(A)=I^*\overline{A}I$ with $I$ and complex conjugation as in Section~\ref{sec-struct}, and the real C$^*$-algebra will then be denoted by $\Aa_I$. Furthermore, only the $K$-group $K_0(\Aa_I)$ will be considered. It is by definition the set of homotopy classes of orthogonal projections (or invertible self-adjoints) in matrix algebras over $\Aa_I$, to which $I$ is extended by $I\otimes \one$. Otherwise stated, $K_0(\Aa_I)$ is the set of homotopy classes of odd symmetric projections in $\Aa$. The reason for picking out this example is that there may be $\ZM_2$ invariants contained in $K_0(\Aa_I)$. It is also possible to study classes of even symmetric invertible self-adjoints (giving classes in $K_0(\Aa_J)$ with $J^2=\one$), or even and odd symmetric skew-adjoints,  and furthermore  classes of unitaries (or even invertibles) which are even or odd real or symmetric, but this full-fledged theory requires various elements of Real $K$-theory and will be developed elsewhere.

\vspace{.2cm}

Topological content will be extracted from a class $[P]_0\in K_0(\Aa_I)$ again by use of Fredholm modules which, however, have a supplementary symmetry property.  An (ungraded version of an) even $\ZM_2$-Fredholm module $(\Hh,F)$ is defined to be an odd symmetric unitary operator $F=I^*F^tI$ such that $[A,F]\in\KM(\Hh)$ for all $A\in\Aa_I$. A pairing will be defined by 
\begin{equation}
\label{eq:even_pairZ_2}
\ll (\Hh,F)\,,\,[P]_0\gg_0
\,\;=\;
\Ind_2(PFP)\;,
\end{equation}
where $P\in\Aa_I$, or matrix algebras thereof. Note that indeed $PFP$ is odd symmetric so that the $\ZM_2$ index is well-defined and due to its homotopy invariance the pairing is independent of the choice of representative. As shown in Theorem~\ref{theo-OddFredFlow} in Section~\ref{sec-oddsympSF},  $\Ind_2(PFP)$ can be calculated as the $\ZM_2$-valued spectral flow of any odd symmetric dilation of $PFP$:
$$
\ll (\Hh,F)\,,\,[P]_0\gg_0
\; \,=\;
\SF_2(2P-\one,F)\;.
$$
Let us now sketch a $K$-theoretic interpretation of this $\ZM_2$ spectral flow by using an adequate $\ZM_2$-mapping cone. It is defined by
$$
\Mm_I
\,=\,
\left\{
(A_s)_{s\in[0,1]}\in C([0,1],\Aa+\KM(\Hh))\,\left|\,I^*\,\overline{A_s}\,I=F^*A_{1-s}F,\;A_s-A_0\in\KM(\Hh),\;A_0\in\Aa_I
\right.\right\}
\;.
$$
This is the fixed point set in $\Mm$ w.r.t. the anti-linear involution  (also denoted by $\tau$)
$$
\tau((A_s)_{s\in[0,1]})
\;=\;
(I^*F^t\overline{A_{1-s}}\,\overline{F}I)_{s\in[0,1]}
\;.
$$ 
Hence $\Mm_I$ is indeed a real C$^*$-algebra. Furthermore, one has an exact sequence of real C$^*$-algebras
$$
0\;\rightarrow\;S_I\KM(\Hh)\;\hookrightarrow\;\Mm_I\;\stackrel{\mbox{\rm\tiny ev}\;}\rightarrow\;\Aa_I\;\rightarrow\,0
\;,
$$
where
$$
S_I\KM(\Hh)
\;=\;
\left\{
(K_s)_{s\in[0,1]}\in C_0((0,1),\KM(\Hh))\,\left|\,I^*\,\overline{K_s}\,I=F^*K_{1-s}F\;
\right.\right\}
\;.
$$
Indeed, for any $A\in\Aa_I$ a lift into $\Mm_I$ is given as before by $\mbox{\rm Lift}(A)_s=(1-s)A+s\,F^*AF$ and this lift can again be chosen self-adjoint for self-adjoint $A$. Based on this construction, one can now transpose the ideas of Section~\ref{sec-mappingcone}.


\end{document}